\documentclass{article}

\usepackage[preprint, nonatbib]{neurips_2023}

\usepackage{tabularx}
\usepackage{paralist}
\usepackage{microtype}
\usepackage{graphicx}
\usepackage{subfigure}
\usepackage{booktabs} 
\usepackage{hyperref}       
\usepackage{microtype}
\usepackage{subfigure}
\usepackage{booktabs} 
\usepackage{url}            
\usepackage{amsfonts}       
\usepackage{nicefrac}       
\usepackage{microtype}      
\usepackage{xcolor}         
\usepackage{xspace} 
\usepackage{amsmath}
\usepackage{amssymb}
\usepackage{multirow}
\usepackage{hhline}
\usepackage{float}
\usepackage{soul}
\usepackage{algorithmic}
\usepackage{algorithm, setspace}
\usepackage{colortbl}
\newtheorem{definition}{Definition}

\usepackage{tikz}
\usepackage{ctable}
\usepackage{pifont}
\usepackage{enumitem}
\usepackage{changepage}

\newcommand{\xmark}{\ding{55}}%

\usepackage[utf8]{inputenc} 
\usepackage[T1]{fontenc}    
\usepackage{hyperref}       
\usepackage{url}            
\usepackage{booktabs}       
\usepackage{amsfonts}       
\usepackage{nicefrac}       
\usepackage{microtype}      
\usepackage{xcolor}         

\usepackage{makecell}

\title{CAPS: A Practical Partition Index for Filtered Similarity Search}

\author{%
 Gaurav Gupta \\
  Rice University\\
  Houston, TX \\
  \texttt{gaurav.gupta@rice.edu} \\
  \And
  Jonah Yi \\
  Rice University \\
  Houston, TX \\
  \texttt{jwy4@rice.edu} \\
  \And
  Benjamin Coleman \\
  Google Inc. \\
  Mountain View, CA \\
  \texttt{benjamin.ray.coleman@gmail.com} \\
  \AND
  Chen Luo \\
 Amazon Search \\
 Palo Alto, CA \\
  \texttt{cheluo@amazon.com} \\
  \And
  Vihan Lakshman \\
  ThirdAI \\
  Houston, TX \\
  \texttt{vihan@thirdai.com} \\
  \And
  Anshumali Shrivastava \\
  Rice University \\
  Houston, TX \\
  \texttt{anshumali@rice.edu} \\
}

\begin{document}

\maketitle

\begin{abstract}
With the surging popularity of approximate near-neighbor search (ANNS), driven by advances in neural representation learning, the ability to serve queries accompanied by a set of constraints has become an area of intense interest. While the community has recently proposed several algorithms for constrained ANNS, almost all of these methods focus on integration with graph-based indexes, the predominant class of algorithms achieving state-of-the-art performance in latency-recall tradeoffs. 
In this work, we take a different approach and focus on developing a constrained ANNS algorithm via space partitioning as opposed to graphs. To that end, we introduce Constrained Approximate Partitioned Search (CAPS), an index for ANNS with filters via space partitions that not only retains the benefits of a partition-based algorithm but also outperforms state-of-the-art graph-based constrained search techniques in recall-latency tradeoffs, with only 10\% of the index size.

\end{abstract}

\section{Introduction}

In recent years, the machine learning community has seen a surge of interest in \emph{vector databases} offering performant similarity search over dense feature representations. Fueled by the breakthrough success of deep neural networks in mapping raw, unstructured data, such as text and images, into semantically coherent embeddings, vector databases now underpin numerous applications in web search \cite{huang2013learning}, product recommendations \cite{nigam2019semantic, li2021embedding, lakshman2021embracing}, social networks \cite{sharma2017hashes}, video analytics \cite{wray2021semantic}, computational biology \cite{engels2021practical} \cite{gupta2021fast}, and many more. 

With the tremendous interest and commercial adoption of similarity search indexes, \emph{Filtered Near Neighbor Search}, which involves retrieving near-neighbor vectors subject to a set of filters, has emerged as an essential feature in many applications. This setting frequently appears in the context of product search, where shoppers often issue queries coupled with constraints such as ``free shipping" or ``under \$10." Various graph-based similarity search algorithms have been adapted in the last few months to provide constraint satisfaction functionality. For example, AIRSHIP \cite{AIRSHIP}, NHQ \cite{NHQ}, and DiskANN \cite{jayaram2019diskann, DiskANNFiltered} all incorporate the filters into the graph walk subroutine that forms the core of their respective search algorithms. However, these approaches all carry limitations (Table \ref{tab:capabilities}) that hinder their practical applicability, such as 
an inability to support a variable number of constraints per query 
or a lack of support for conjunctive ANDs (e.g. ``free shipping AND under \$10").

While these aforementioned prior works in Filtered Near Neighbor Search focus on integrating with popular graph-traversal algorithms, many industry-scale similarity search databases still use space partition-based approaches due to their unique advantages \cite{FAISS,guo2020accelerating}. For example, graph-based ANNS algorithms can incur frequent cache misses due to random node access \cite{coleman2022graph} and cannot easily be parallelized~\cite{HNSW}. Conversely, partition-based algorithms are trivial to parallelize and have a cache and accelerator-friendly data access pattern. To our knowledge, however, there is currently no efficient algorithm or structure for constrained search over partitions short of filtered brute force.

\begin{table}[t]
\centering
\label{tab:capabilities}
\resizebox{\linewidth}{!}{
\fontsize{7}{7.5}\selectfont
\begin{tabular}{|c|c|c|c|c|}
  \hline
  Algorithm & Variable Constraints & Conjunctive ANDs & Sparse  Attributes & Dynamic Insertions \\
  \hline
  NHQ \cite{NHQ} & \cellcolor{red!25}\xmark & \cellcolor{green!25}\checkmark  & \cellcolor{red!25}\xmark & \cellcolor{red!25}\xmark \\
  AIRSHIP \cite{AIRSHIP} & \cellcolor{green!25}\checkmark & \centering{\cellcolor{green!25}\checkmark} & \cellcolor{red!25}\xmark & \cellcolor{red!25}\xmark \\
  DiskANN \cite{DiskANNFiltered} & \cellcolor{green!25}\checkmark & \centering{\cellcolor{red!25}\xmark} & \cellcolor{green!25}\checkmark & \cellcolor{green!25}\checkmark \\
  CAPS (ours) & \cellcolor{green!25}\checkmark & \centering{\cellcolor{green!25}\checkmark} & \cellcolor{green!25}\checkmark & \cellcolor{green!25}\checkmark \\
  \hline
\end{tabular}
}
\caption{Summary of capabilities of open-source Filtered Near Neighbor Search algorithms.}
\vspace{-4mm}
\end{table}

Motivated by these challenges, we propose CAPS (Constrained Approximate Partitioned Search), a new filtered search algorithm based on a hierarchical partitioning that guarantees that the returned set of vectors matches all constraints while demonstrating improved efficiency in the high-recall regime.

\subsection{Contributions}
We begin with an exploratory experiment that sheds light on the nature of filtered near-neighbor search.
Additionally, we develop methods to address uncorrelated attributes and conjunctive predicates (i.e. AND constraints). We make the following specific contributions.

\textbf{Constrained Space Partitioning:} To our knowledge, we present the first non-trivial approximate algorithm for filtered near neighbor search via space partitions. One of the key problems with filtered search in a partitioned index is that most indexes have large partitions that may only sparsely satisfy the constraint. To address this, we introduce a Huffman tree-inspired hierarchical sub-partitioning algorithm that shards the data based on vector embeddings followed by attributes, which we call an \emph{Attribute Frequency Tree} (AFT). We also experiment with a learned strategy for jointly partitioning the embedding and attributes, which achieves superior results on certain benchmarks.

\textbf{Practical Utility:} We present, to our knowledge, the first efficient and practically useful constrained search algorithm that supports a varying number of attributes per query, conjunctive ANDs, support for rare attributes, and dynamic insertions. Our proposed index also uses a fraction of the space compared to previously proposed graph-based filtered search algorithms. 

\textbf{Real-world Evaluation:} Prior work in this area relies on synthetic tasks constructed from standard near-neighbor benchmarks with varying distributional assumptions on the attributes (e.g. clustered or random). We apply our methods to a large-scale industrial e-commerce application and find that real-world constraints are power-law distributed. We use the power law structure to enhance our sub-partition tree algorithm and further improve performance.\\

\section{Filtered Near Neighbor Search}
\label{sec:definition}
Modern search applications require simultaneous similarity search of vector embedding data and filtering by categorical attributes. We call this framework \textit{filtered near neighbor search} because it combines ideas from similarity search with more traditional relational database operations.\\
\textbf{Attribute Model:} We adopt the industry standard attribute model, where an attribute is a tag for an item that can take multiple categorical values. An item can have any number of attributes, each with one value of an attribute. We begin with a formal definition of Query Filter and Filtered Search.
\begin{definition}
\label{def:QueryConstraint}
Query Filter\\
A \emph{query filter} is a function $C: \mathbb{N}^L \times \mathbb{N}^L \to \{\mathrm{True}, \mathrm{False}\}$ such that, given a data attribute vector $a \in \mathbb{N}^L$ and a query attribute vector $b \in \mathbb{N}^L$, $C(a, b) = \mathrm{True}$ if $a[i] = b[i] ~ \forall i$ and $\mathrm{False}$ otherwise. 

\end{definition}

The near-neighbor search has many theoretical formulations, leading to slightly different problem statements. For example, the $(R,c)$-approximate near neighbor problem asks that we retrieve all points within radius $R$ of the query and none other (\cite{indyk1998approximate,datar2004locality}). The $k$-nearest neighbor problem~\cite{coleman2020sub} is to identify all of the $k$ closest points to a query with high probability. We state the constrained problem in a way that is agnostic to the specific choice of the near-neighbor formulation.

\begin{definition}
\label{def:constrained_knn}
Filtered Near-Neighbor Search \\
Given a dataset $D = \{(x_1, a_1), \dots (x_N, a_N)\}$ of embedding vectors $x \in \mathbb{R}^d$ and associated attributes $a\in \mathbb{N}^L$, a query $q$ and a query filter $C(q, b)$, solve the near neighbor problem of interest on the \textit{restricted dataset} $D_C = \{(x_i,a_i) \in D: C(b, a_i)\}$.
\end{definition}

Definition~\ref{def:QueryConstraint} defines the query filter constraint, and Definition~\ref{def:constrained_knn} defines the constrained similarity search problem as a \textit{filtered} version of the standard similarity search. This is well-motivated by industrial applications, where practitioners often wish to perform a query with the SQL-style predicate ``\texttt{WHERE} $dist(x,q) < R$ \texttt{AND} $x \in D_C$.'' 

\section{Challenges of Filtered Search}
In this section, we seek to understand the conditions under which existing constrained search strategies fail to retrieve results in a performant manner.

There is a fundamental tension between satisfying the constraints and taking the optimal path through the dataset for similarity search. While this could, in principle, be solved by constructing an index for each possible attribute (or by incorporating the attribute into the vector embedding), such solutions are insufficient in practice for two reasons. 

\begin{itemize}[align=left, leftmargin=*]
  \item Often, the attributes are either unknown or of high cardinality, precluding algorithms that pre-index the data. In other words, $D_C$ must be identified \textit{dynamically} and on a query-dependent basis.
\item Elements that do not satisfy the constraint must often be examined en-route to the solution.
For example, it is not straightforward to choose between a slightly-further partition with many items in $D_C$ and a slightly-closer partition with few valid items. 
\end{itemize}

Ultimately, all constrained search approaches must find a way to integrate the searching and filtering operations efficiently. Below, we summarize the two possible approaches. 

\textbf{Search, then filter:} A natural solution is to post-filter the items that are identified by the similarity search index. This approach works well when many items in the dataset satisfy the constraint, because then it is likely that only a few extra results will need to be returned from the index. However, it is problematic when $D_C$ is sparsely populated.

Sparsity is a particularly serious issue in high dimensions, because valid candidates may be spread over an exponentially large space (in $d$). For graph-based algorithms, this manifests as an exploration set that grows rapidly due with the fan-out of each node \cite{malkov2018efficient}. For partition-based algorithms, this manifests as low recall unless a large number of clusters are probed \cite{FAISS}.

\textbf{Filter, then search:} Another equally-viable approach is to apply an attribute-based filter \textit{before} conducting a brute force similarity search over any items that match. This approach has the exact opposite sparsity properties: highly sparse queries are very easy, because then we need only explore a few items, but such a brute force search will be prohibitively expensive if many items satisfy the constraints. 

\subsection{The unhappy middle} 
Extremely sparse constraints can be solved quickly by cascading an inverted index (or randomized approximation) with a brute-force search. Extremely dense constraints can be solved quickly by search-then-filter. Therefore, the only algorithmically interesting regime is when there are too many items to brute force $D_C$ but too few to post-filter a standard similarity search. We call this region \textit{the unhappy middle}. Unfortunately, most real-world queries fall into this regime. To illustrate this problem, we conducted an experiment using a partition-based search over MNIST \cite{lecun1998mnist} with synthetic attributes. We randomly distribute the attributes i.i.d. throughout the whole dataset with different probabilities, then query while constraining on each attribute. We adjust the fraction of elements that satisfy the constraint (the sparsity) to investigate the effects of sparsity on search.

\begin{figure}[h]
\centering
  \includegraphics[width=3.7in]{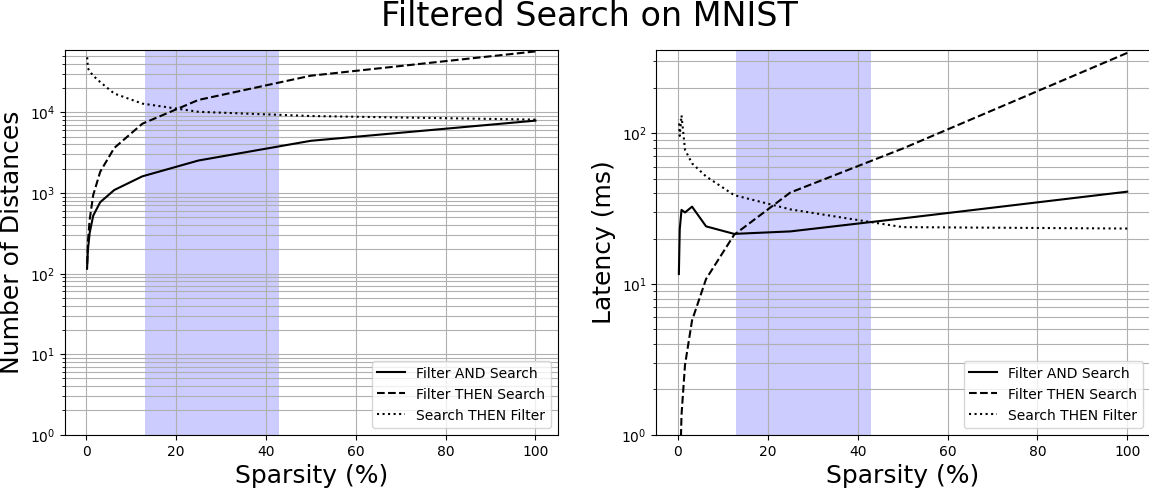}
  \caption{Distance calculations (left) and latency (right) for filtered search on MNIST , subject to recall > 95\%. For sparse attributes, it is optimal to filter first, then search. For dense attributes, it is sufficient to filter the search output. 
  }
  \label{fig:mnist_experiment}
\end{figure}

Figure~\ref{fig:mnist_experiment} shows the effect of sparsity on the number of items examined (left) and the overall search latency (right). The results confirm our intuition: highly sparse and highly dense constraints can be adequately addressed with separate search and filtering. However, there is a central region (highlighted) where it is expensive to brute-force the entire set of valid results and equally expensive to find enough search results for post-filtering \footnote{We note that the latency numbers  (Figure~\ref{fig:mnist_experiment}) are somewhat artificial, as we  used a simplified and unoptimized version of our algorithm using Python. Nonetheless, we believe this offers valuable insight into the nature of constrained search.} 

We acknowledge that the differentiation between pre-filtering and post-filtering approaches has been addressed in prior discussions, including the work by Pinecone \cite{pinecone}. However, our contribution lies in the empirical analysis of the unhappy middle problem, which, to the best of our knowledge, has not been previously characterized. This analysis highlights the importance of developing improved algorithms in order to make further advancements in hybrid query processing.

\section{Related Work}
\label{sec:RelatedWork}

Approximate solutions to the near-neighbor search problem exhibit tremendous algorithmic diversity. At various times over the last 30 years, locality-sensitive hashing ~\cite{andoni2008near, wang2017flash}, quantization and partitioning methods \cite{jegou2010product, dong2019learning, guo2020accelerating, gupta2022bliss, engels2021practical}, trees~\cite{ram2019revisiting, beygelzimer2006cover, izbicki2015faster}, and graph-based methods \cite{HNSW, iwasaki2018optimization} have each represented state of the art.  
Out of this, graphs and partitions have emerged as the two dominant data structures.

\textbf{Partitions:} A large family of successful similarity search algorithms, such as FAISS-IVF \cite{FAISS}, splits the dataset into partitions/clusters. This improves the search latency because the query does not need to examine all data points but only a few clusters that are likely to contain neighbors. The search algorithm identifies a small subset of nearby clusters, typically via a near-neighbor search on cluster identifiers. This follows the brute force near neighbor search over the points in clusters.

\textbf{Graphs:} Graph-based methods have traded positions with partition-based methods at the top of the leaderboard for several years~\cite{AnnBenchmark}. Graph algorithms locate near neighbors by walking the edges of a graph where each point is (approximately) connected to its $k$ nearest neighbors. This area has focused on improving graph properties using diversification, pruning, hierarchical structures, and other heuristics~\cite{malkov2018efficient}. Recent work in this area has focused on systems properties of graph implementations~\cite{coleman2022graph}, dimension reduction and theoretical analysis~\cite{prokhorenkova2020graph}, and selective, pruned computation~\cite{chen2023finger}. It is also common for a graph index to serve as the partition identification step of a partitioned index (e.g. Neural LSH \cite{nlsh} and FAISS with HNSW pre-indexing).

\subsection{Hybrid query search}
The state of the art in constrained search has recently moved beyond simple compositions of filters and search algorithms. A variety of \textit{hybrid} approaches have been proposed to integrate the constraint satisfaction into the search index. In this section, we outline several recent approaches.

\textbf{AIRSHIP:}
AIRSHIP is a constrained near-neighbor algorithm based on the HNSW near-neighbor graph. 
Most graph algorithms employ a variation of beam search to walk through the graph and identify neighbors. Beginning from a randomly selected set of initial nodes, beam search maintains a list of ``exploration candidates.'' The best candidates (those closest to the query) are retained during the search as a tentative set of near-neighbors. Once the algorithm has explored a sufficiently large fraction of the graph and exhausted its supply of candidates, it returns the best candidates as the neighbors.

\textbf{DiskANN:} Originally introduced in 2019, DiskANN \cite{jayaram2019diskann} is a graph-based search index capable of indexing and searching over billion-scale vector collections with only 64GB of RAM. The authors subsequently built on this work by introducing capabilities for real-time updates \cite{singh2021freshdiskann} and support for filtered queries \cite{DiskANNFiltered}. The key insight behind the filtered DiskANN family of algorithms is to incorporate knowledge of the attributes into the graph indexing stage so that the greedy search can leverage this information. However, this algorithm only supports the simplified case of a single constraint per query. While a single filter suffices to support disjunctive ORs (via a union of independent searches), the authors note that supporting conjunctive ANDs would require new ideas.  

\textbf{NHQ}: Native Hybrid Query \cite{NHQ} is also a navigable proximity graph-based constrained near-neighbor search algorithm. It builds a graph like HNSW, using a fused distance metric given by $d_f(.) = d_v(.) + wd_a(.)$, where the distance between any two points is given by the weighted summation of Euclidean distance between corresponding vectors and hamming distance between the binary attribute vector. 

\textbf{AnalyticDB-V:} AnalyticDB \cite{analyticdb} seeks to provide support for SQL-style queries over data that consists of both relational tables and embeddings of unstructured entities. The core insight behind this work is to develop four query plans and an associated cost optimizer to efficiently determine how to order the filtering and search operations. These query plans include brute force search, product quantization bitmap scans, and graph-based searching. We do not evaluate this method in our experiments due to a lack of publicly available code. Still, we note that our contributions complement AnalyticDB in that CAPS can be viewed as a potential query plan within this larger database engine. 

\section{CAPS: Interleaved Filtering and Search}
\label{sec:CAPS}
Our proposed method allows for simultaneous filtering and search within a clustered near-neighbor index. We are looking for an algorithm that performs equally well in the dense-attribute regime (the head), the sparse regime (the tail), and the unhappy middle. To enable constrained search, we modify the clustering portion of the algorithm and augment the cluster identification process with constraint checks. The intuition for this direction is as follows.

\begin{figure}[h]
  \includegraphics[scale=0.35]{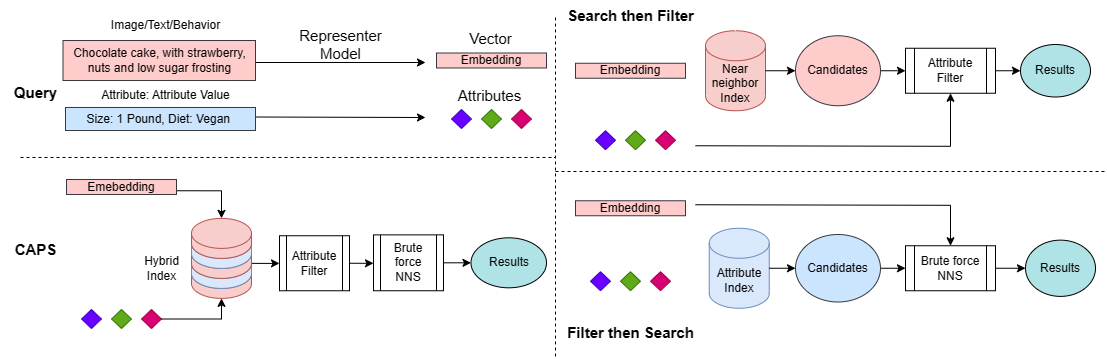}
  \caption{
  Filters may be added to similarity search as a pre-filter (search then filter) or a post-filter (filter then search).
  We propose a hybrid search for constrained queries: CAPS, which performs an interleaved series of filter and partition lookup operations.
  }
  \label{fig:vanillaApproach}
\end{figure}

\subsection{Intuition}
Consider a standard partitioned index, where we identify partitions likely to contain nearby points.
Once a suitable set of partitions have been found, the neighbors which satisfy the constraints can be found via brute-force attribute matching followed by the metric distance computation on the positive attribute matches. If we examine $m$ different partitions $\{P_1, ... P_m\}$, then we will require a total of $\sum_{bin=1}^{m}|P_{bin}|$ brute-force attribute matches and distance computations. 

We can apply filtering within the partition to reduce this computational burden, but this still requires passing every point in the partition through the filter. CAPS breaks this limitation by sub-partitioning each cluster (at indexing time) based only on the attributes. By layering a similarity search index with an attribute index, we effectively perform a \textit{filter-then-search} step after doing a preliminary \textit{search-then-filter} over the centroids.

Our sub-partition construction algorithm is motivated by two observations. First, the sub-partitions should be non-overlapping. Second, we should refuse to search a sub-partition if none of its contents satisfy the constraint.\footnote{Equivalently, if \textit{any} point in a sub-partition is valid, we \textit{should} search the sub-partition.}
The first observation manifests as a constraint in the sub-partitioning algorithm: for each cluster $P_{bin}$, $P_{bin}= \bigcup_{j=1}^{h+1}p_{bin,j}$ and $p_{bin,j} \cap p_{bin,i} =\emptyset$ for $j \neq i$. In other words, each cluster has $h+1$ disjoint sub-partitions. 

\begin{figure*}[ht]
  \includegraphics[scale=0.36]{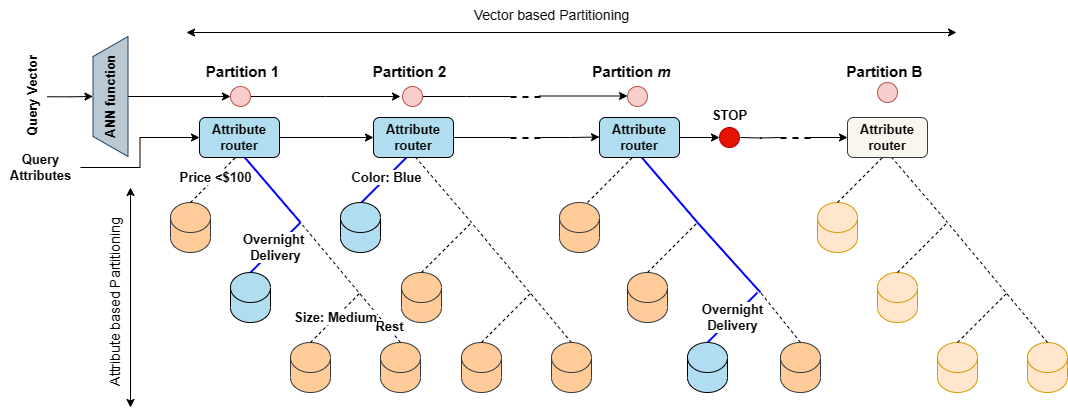}
  \caption{The query $q$, selects the closest $m$ partitions and corresponding sub-partitions further on the truncated AFT. The $h$ sub-partitions represent the disjoint sets of high occurrence attributes, and the last sub-partition contains the remaining points. }
  \vspace{-2mm}
  \label{fig:query}
\end{figure*}

Our second observation means one can check all sub-partitions for query filter match. Instead, we can quickly identify relevant sub-partitions in $O(1)$ time. This can be done if
we cluster points with one common attribute value into the same sub-partitions so that only a few sub-partitions are likely to be valid.
This is achieved by organizing the points using a frequency-based partitioning over the attributes. We explain this partitioning scheme in the following section and in Figure \ref{fig:query}. This reduces the query complexity to $\sum_{bin=1}^{m}\sum_{j=1}^{h+1}C(b,p,bin,j)|p_{bin,j}|$, where $C(b,p,bin,j) = \mathbb{I}(b=A(p_{bin,j}))$, which checks whether the sub-partition contains the query attribute $b$. 

\begin{minipage}{0.55\textwidth}
\begin{algorithm}[H]
  \begin{algorithmic}
   \STATE {\bfseries Input:} Partitioning Model $f(.)$, dataset $D ={(x_i,a_i)}$\\
   \STATE Learn partitioning Model $f(.)$ = $\mathcal{L(D)}$\\
    \FOR{ $i \in 1..N$}
        \STATE $bin = f(x_i, a_i)$, \ \ $P_{bin} \leftarrow (i)$\\
    \ENDFOR
    \FOR{ $bin \in 1..B$} 
        \STATE $S = \phi $\\
        \FOR{ $j \in 1..h$}
            \STATE $t$ = top-attribute$(P_{bin} - S)$\\
            \STATE $p_j  \leftarrow (i), t \in a_i, i \in (P_{bin} - S)$  \\
            \STATE $A(p_j) = t$   \#attribute tag\\ 
            \STATE $S = S \cup p_j$
        \ENDFOR
        \STATE $p_{h+1}  \leftarrow (P_{b} - S)$  \\
    \ENDFOR
    \STATE {\bfseries Output:} Hierarchical partitions $P$, $p$
   \end{algorithmic}
     \caption{CAPS Index construction }
   \label{alg:IndexConstructionAlgo}
\end{algorithm}

\end{minipage}
\begin{minipage}{0.42\textwidth}

\begin{algorithm}[H]
  \begin{algorithmic}
   \STATE {\bfseries Input:} Partitioning Model $f(.)$\\ \ \ \ \ \ \ \ \ \ \ \ \ dataset $D ={(x_i,a_i)}$\;
    \STATE {\bfseries Input:} Query point $(q,b) \in Q$\;
    \STATE $\mathcal{R}(q) = \phi$
    \STATE \# Select top $m$ partitions
   \STATE $\{bin_1,..bin_m\} = f(q)$ \\
    \FOR{ $bin \in \{bin_1,..bin_m\}$ }
        \STATE IF $A(p_j) = b$ for any $j \in 1..h$
        \STATE \ \ \ $\mathcal{R}(q) = \mathcal{R}(q) \cup p_j$ 
        \STATE ELSE
        \STATE \ \ \ $\mathcal{R}(q) = \mathcal{R}(q) \cup p_{h+1}$ 
        \STATE candidate set for $q$ $ =\mathcal{R}(q)$
    \ENDFOR 
    \STATE ${R(q)}$ = Attribute Match$(\mathcal{R}(q),q) $\\
    \STATE Result = topk Vector Match $({R(q)})$
    \STATE {\bfseries Output:} Result\\
   \end{algorithmic}
     \caption{CAPS Query }
   \label{alg:IndexQueryAlgo}
\end{algorithm}
\end{minipage}

\subsection{Index creation}
At index time, we perform two-level partitioning of the data $D$. The first level is done by applying any existing partitioning method on the embedding vectors. Balanced Kmeans clustering from FAISS-IVF \cite{FAISS} is a common choice in partitioned indices. This partitioning does not require access to the attribute information, but instead clusters the dense vectors using a similarity measure such as the Euclidean distance, Cosine distance, or Inner-product. 

However, it is also possible to integrate the attributes into the partitioning. This can improve performance if we have reason to suspect that the attributes correlate or cluster with the embeddings. While certain algorithms, such as NHQ, explicitly fuse the attribute and vector similarity measures into a single distance metric, this is highly dataset-dependent and is difficult to tune. This is especially problematic in production settings, where some -- but not all -- attributes are correlated with embeddings. 

Recent work on learned partitioning provides a nice avenue to integrate attributes into the first-level clustering. For example, the BLISS method \cite{gupta2022bliss} uses an iterative learning scheme to learn the partitions for any binary \textit{relevant-vs-non-relevant} similarity match oracle. Therefore, we can use the true filtered near neighbors of a set of points in the dataset to learn the BLISS partitions. We consider two different setups. In CAPS-BLISS1, we do not use the attribute or filtering information, but instead train and index partitions based only on the vector near neighbors' labels. In CAPS-BLISS2, we use the filtered near neighbors labels against the data vectors.

We will use the function $f(\cdot)$ to denote the first-level partition function that assigns a partition to an input point. By applying $f(\cdot)$ to the elements of the dataset, we get the first-level partitions. The next step is to construct second-level sub-partitions that shard by attribute. We create these partitions by exploiting the power law distribution of the attributes, using a hierarchy called an , which we call an \textit{Attribute Frequency Tree} (AFT). A given partition $P$ is split with the points satisfying the most frequent attribute value, and the remaining set is split further with the same strategy. This creates a tree where each level has two disjoint sub-partitions, with one having the selected attribute value. In the interest of lower index memory, we truncate the tree after a fixed number of levels denoted by the height $h$. This truncation scheme is particularly well-suited to attributes that follow a power law distribution. We refer to Algorithm \ref{alg:IndexConstructionAlgo} and Section \ref{sec: ablation} for a detailed analysis of the tree depth $h$.

\subsection{Query Algorithm}
The query algorithm uses the learned partitioning function $f(\cdot)$ and selects the top $m$ similar partitions, where $m$ is the number of partitions to probe. For the search to be efficient, we choose $m \ll B$. The query proceeds by selecting sub-partitions within each of the $m$ partitions, using the truncated AFT. Each level of the tree splits into a leaf and a sub-tree, where the split is performed by an attribute that is common to all items in the leaf (an \textit{identifier}); thus we have $O(h)$ identifiers. We hash the positions of the identifiers so that the process of identifying sub-partitions in each first-level partition has $O(1)$ complexity. To get the final candidate set for brute force, we join all of the valid sub-partitions into a small candidate set that (mostly) contains valid points. In the setting where we constrain on multiple attributes (e.g., an \texttt{OR} constraint), multiple sub-partitions from each partition are included in the join.

\section{Evaluation}
\label{sec:evaluation}
\textbf{Baselines: }
We use compare CAPS with the existing SOTA algorithms for hybrid search. This includes NHQ \cite{NHQ}, AIRSHIP \cite{AIRSHIP}, Filtered DiskANN's filtered index (FI) and Stiched Index (SI) \cite{DiskANNFiltered}. DiskANN is limited to working only with one attribute per query; hence we evaluate DiskANN with only one attribute per query (even though multiple attributes are specified). Where possible, we use the original authors' implementations. 

\textbf{Experiment setup:}
We measure the recall of the top 100 filtered neighbors (following definition \ref{def:QueryConstraint}) against the filtered search index output. The Recall$K$@$K$ is the intersection of the top$K$ items retrieved from the index with the true top$K$ items. 
Indexing performed using 64 threads and we report single-thread query results (as is standard for kNN experiments~\cite{AnnBenchmark}).

\begin{figure*}
  \includegraphics[scale=0.35]{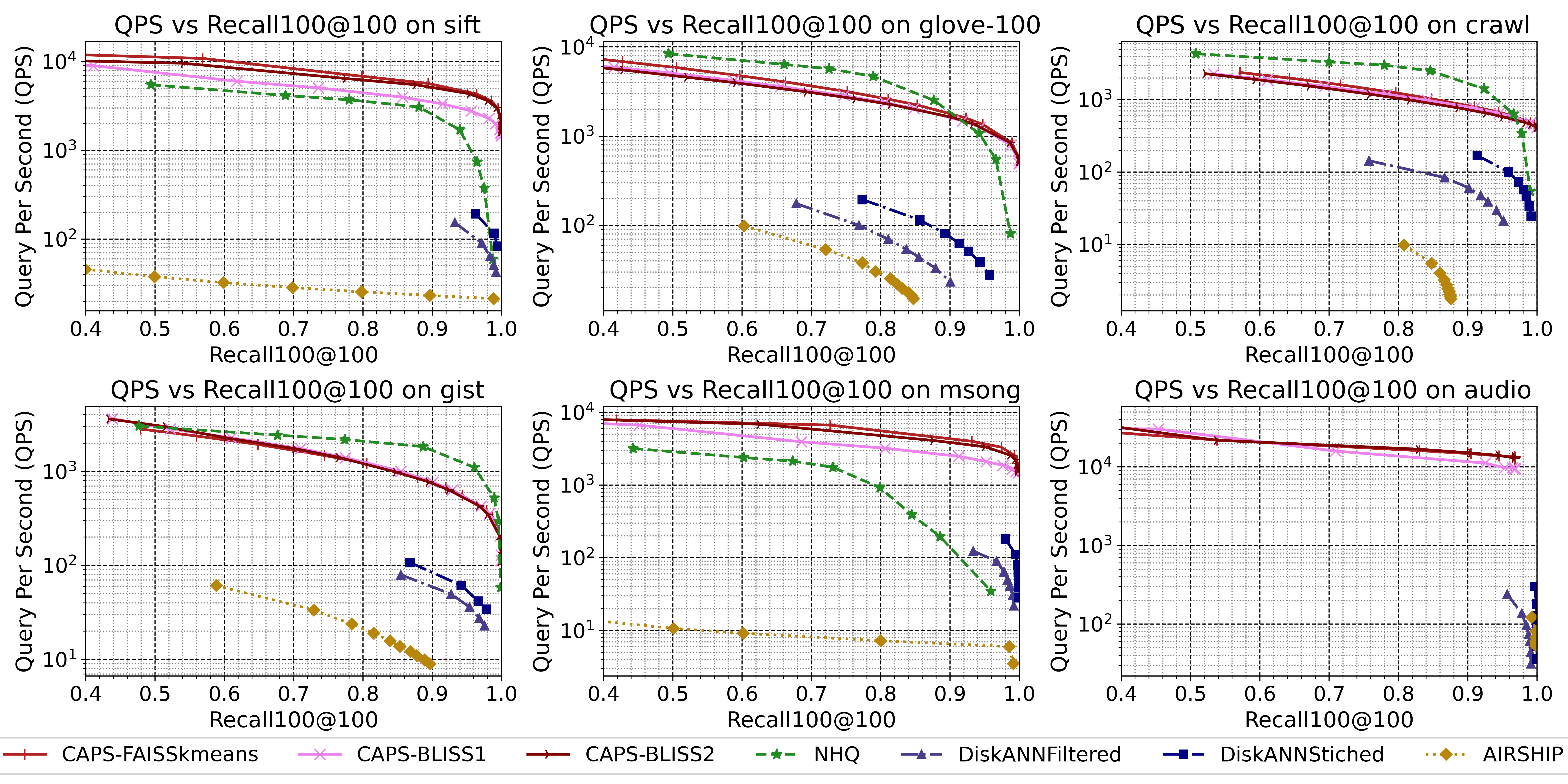}
  \vspace{-1mm}
  \caption{Recall100@100 vs Query per second tradeoff on public datasets (top-right is best).}
  \label{fig:RecallVsQPS_7Attri}
\end{figure*}

\textbf{Datasets:}
We use six publicly available real-world near-neighbor datasets SIFT, Glove-100, GIST, Crawl, Audio, and Msong. Because the community currently lacks any open constrained search benchmark datasets, we randomly generate attributes for each dataset with $L=3$ following the same procedure as \cite{NHQ}. Each attribute can have multiple categorical values and is distributed with exponential distribution motivated by a real-world scenario in Figure \ref{fig:AmazonattributeDist}. 
\textbf{Discussion:} 
Figure \ref{fig:RecallVsQPS_7Attri} compares CAPS with FAISSkmeans, BLISS1 and BLISS2 with baselines on Recall100@100 against QPS. AIRSHIP suffers from the sparsity of the constraints highly for all datasets apart from SIFT. DiskANN achieves high recall but it has slower latency on a single thread. NHQ is competitive, but CAPS excels mostly in the high recall region. We also note that NHQ contains an attribute-embedding fusion hyperparameter that is challenging to tune well. Table \ref{tab:ablation1} presents the overhead added by the indexing structure (e.g., graph or hierarchical partition). We achieve a very low index size with 10x smaller overhead than the next-cheapest baseline (NHQ). Please refer to the supplementary section for the dataset and code details.

\begin{table}[t]
\centering
\caption{Indexing overhead (MB) and Construction time (Sec.) of CAPS and against the baselines.}
\label{tab:ablation1}
\resizebox{\linewidth}{!}{
\fontsize{7}{7.5}\selectfont
\begin{tabular}{l l l l l l l l l l l l l }
\toprule
Data & \multicolumn{2}{c}{NHQ} & \multicolumn{2}{c}{AIRSHIP} & \multicolumn{2}{c}{DiskANN-FI} & \multicolumn{2}{c}{DiskANN-SI} & \multicolumn{2}{c}{CAPS-FAISSkm} & \multicolumn{2}{c}{CAPS-Bliss} \\
\cmidrule(lr){2-3} \cmidrule(lr){4-5} \cmidrule(lr){6-7} \cmidrule(lr){8-9} \cmidrule(lr){10-11} \cmidrule(lr){12-13}
& Size & Time & Size & Time & Size & Time & Size & Time & Size & Time & Size & Time \\
\midrule
SIFT  & 75   & \textbf{26}  & 60    & 218   & 357   & 346.6   & 320.4   & 292.2   & \textbf{8.2}    & 69.4    & \textbf{8.8}    & 129 \\
Glove & 34   & 131.8 & 72   & 255   & 254   & 371.11  & 142.8   & 205.5   & \textbf{9.5}    & \textbf{86.8}    & \textbf{11}     & 177 \\
GIST  & 43   & \textbf{47.1 } & 90.4 & 899   & 295.2 & 565.5   & 192.8   & 848.8   & \textbf{12}     & 150.6   & \textbf{9.7}    & 171 \\
Crawl & 67   & \textbf{168}   & 79.4 & 850   & 591.4 & 738.9   & 284.2   & 611.1   & \textbf{18}     & 377.7   & \textbf{18}     & 415 \\
Audio & 3.3  & \textbf{2.83}  & 3.38 & 7.1   & 17.38 & 32.2    & 10.38   & 37.8    & \textbf{0.62}   & 12.6    & \textbf{0.86}   & 8.6 \\
Msong & 73   & \textbf{78}    & 90.4 & 551   & 397.6 & 373.3   & 192.8   & 360     & \textbf{9.3}    & 118.3   & \textbf{9.1}    & 147 \\
\bottomrule
\end{tabular}
}
\end{table}

\subsection{Ablation Study}
\label{sec: ablation} 
\begin{figure*}
  \includegraphics[scale=0.35]{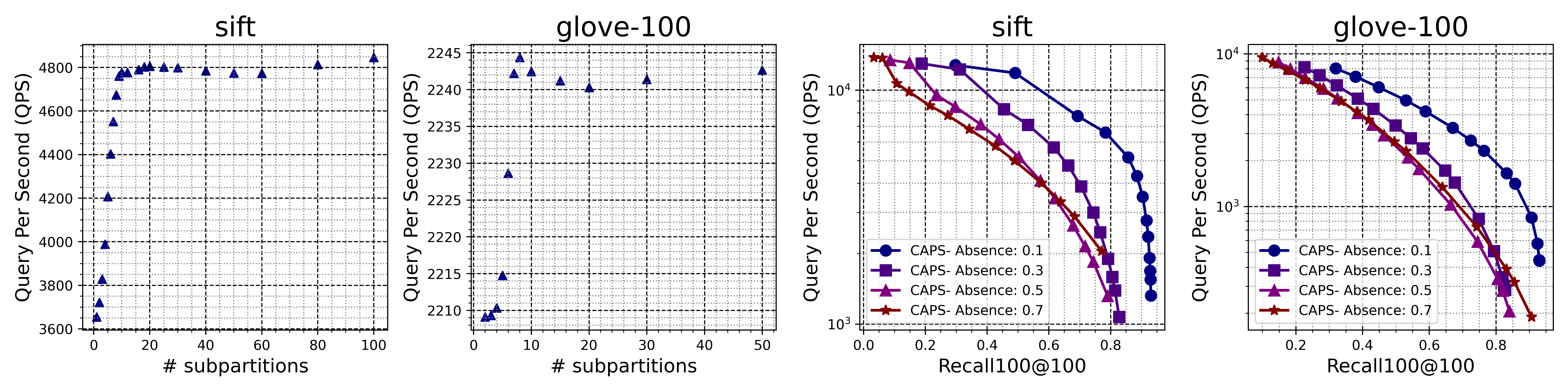}
  \vspace{-2mm}
  \caption{(1-2) The query per second (QPS) increases with an increase in the number of sub-partitions ($h+1$). For each experiment, Recall values remain unchanged. (3-4) QPS vs Recall for a variable number of query attributes. A higher absence fraction implies a fewer number of attributes per query.}
  \label{fig:ablationplots}
  \vspace{-2mm}
\end{figure*}

\textbf{Attribute frequency tree height ($h$):} We evaluate CAPS-FAISSKmeans with a varying number of subpartitions ($h+1$), which are formed at each level of the truncated attribute frequency tree. These subpartitions are non-overlapping groups that can be easily identified using a specific attribute. Consequently, retrieving a single subpartition from each partition can be done in constant time during a query.
As the size of the subpartitions becomes smaller, the computational burden associated with brute force query filtering reduces, in effect reducing the query latency (Figure \ref{fig:ablationplots}).

\textbf{Varying number of attributes ($L$):}
We also vary the number of attributes $L$ involved with the query. In other words, any missing attribute in the query might lead to probing more than one sub-partition from a partition. Figure \ref{fig:ablationplots} (1-2) shows an absence fraction representing the probability of an attribute value missing from the query. Our base set of experiments (Figure \ref{fig:RecallVsQPS_7Attri}) represents absence 0. On the other side, absence 1 will denote only vector-based near-neighbor search.

\subsection{Case Study with Amazon}
\label{sec:amazon}

To evaluate CAPS in a real-world setting, we conducted a case study with the Amazon.com search auto-completion service, which provides customers with search suggestions given a partially typed query. This service requires a low-latency near-neighbor search algorithm to find the corresponding candidate queries. In addition, the suggested query needs to meet certain search constraints, such as preserving the same product type and brand as the original query.

\begin{figure}[htbp]
  \begin{minipage}[b]{0.6\linewidth}
    \centering
    \includegraphics[width=0.65\linewidth]{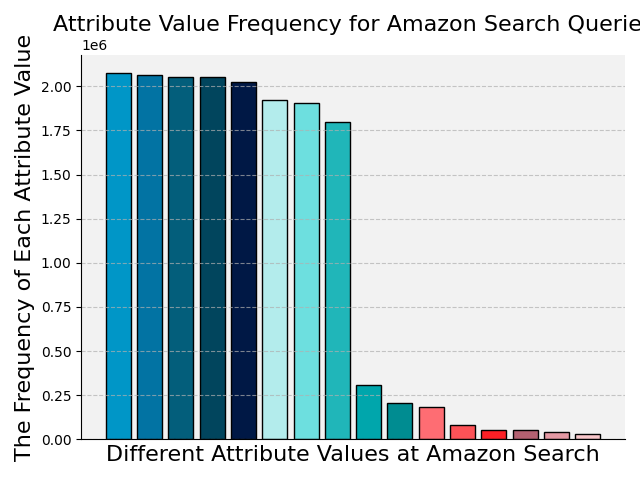}
    \label{fig:AmazonattributeDist}
  \end{minipage}%
  \begin{minipage}[b]{0.4\linewidth}
    \centering
    \resizebox{\linewidth}{!}{
    \fontsize{8}{8}\selectfont
      \begin{tabular}{lccc}
        \toprule
        Methods        & \makecell[l]{Query Per Sec \\(QPS) ($\uparrow$)} & R@100 ($\uparrow$) \\
        \midrule
        Production     & 1.0        & 1.0        \\
        \midrule
        \makecell[l]{CAPS }       & \textbf{5.56}          &\textbf{ 1.2 }         \\
        \bottomrule
      \end{tabular}
      }
    \label{tab:amazon}
    \vspace{1cm}
  \end{minipage}
  \vspace{-4mm}
    \caption{\textbf{Left:} Distribution of attribute values for Amazon search queries. \textbf{Right:} Comparison between CAPS and the production system in Amazon. Due to the confidential nature of the metrics, we report the values relative to production}
    \vspace{-4mm}
\end{figure}

In this case study, we utilize the production model to generate the keyword's embedding \cite{jiang2022short} and extract possible attributes\footnote{We use a sample of 11 possible binary attributes for this case study.} The frequency distribution of attribute values is presented in Table \ref{fig:AmazonattributeDist}, and the dataset dimensions are $N = 8$M and $d = 768$. It is evident that the Amazon attribute distribution follows a power-law distribution pattern with a number of rare constraints. We compared our method with the existing approximate near-neighbor search algorithm in production. As shown in Table \ref{tab:amazon}, CAPS outperforms the production algorithm in terms of query time and recall.

\section{Conclusion}
In this work, we present CAPS, a practical algorithm for filtered approximate near-neighbor search that achieves state-of-the-art performance in the high-recall regime. While the majority of previously proposed filtered search algorithms focus on integrating with graph-based traversal, CAPS provides a novel solution via jointly partitioning the space of embeddings and attributes. This partitioning achieves a smaller space footprint than graph indexes while also supporting critical practical features such as a variable number of query attributes, conjunctive constraints, and dynamic insertions. We also find that the hierarchical partitioning scheme at the core of CAPS is particularly well-suited for power law-distributed attributes, which we validate in a case study with the Amazon.com product search engine. These results demonstrate the effectiveness of space partitioning methods in the filtered search setting and provide practitioners with an immediately useful tool for vector search with constraints.

\bibliographystyle{plain}
\bibliography{references}

\newpage
\section{Appendix}

\subsection{Additional experiment details}
\label{sec:addExpDetails}
\paragraph{Metric}
The metric in use is defined by Recall$K$@$K$ = $\frac{|\text{Index-top}K \ \cap \ \text{True-top}K|}{K}$. Here the $K$ is the number of generated outputs and the ground-truth true constrained near neighbors. 

\paragraph{Hardware:}
We use a Ubuntu 20.04 machine equipped with 2 AMD EPYC 7742 64-core processors and 1.48TB of RAM. All experiments are performed in RAM. CAPS is written in C++ and compiled using the GNU Compiler. Refer to the anonymous GitHub link\footnote{https://github.com/gaurav16gupta/constrainedANN} for the code.\\

\paragraph{Groundtruth:}
\label{bruteForceCAPS}
For the ground truth top$K$ constrained near neighbors, we perform a (slow) exact search over the set of valid points for each query,
i.e., the ground truth search results match the query attributes exactly and are the near neighbors in $D_C$ (see Definition 1).

\subsection{Notations}
\begin{table}[h]
\centering
\caption{Notation Used in the Paper}

\label{tab:notations}
\resizebox{\linewidth}{!}{
\begin{tabular}{ l l l l l l}
\toprule
Symbol & Description & Symbol & Description & Symbol & Description\\
\midrule
$D$ & Dataset & $x$ & data point vector & $a$ & data point attributes\\
$L$ & number of attributes & $q$  & query & $b$ & query attribute \\
$C(.)$ & constraint function & $N$ & Number of points & $f(.)$ & assignment function\\
$P$ & Partition & $p$ & sub-partition & $B$ & number of partitions\\
$h$ & Huffman tree height & $A(.)$ & sub-partition attribute & $m$ & num cluster to probe\\
\bottomrule
\end{tabular}
}
\end{table}

\subsection{Dataset details}
We use six publicly available real-world near-neighbor datasets SIFT, Glove-100, GIST, Crawl, Audio, and Msong. 

\begin{table}[h]
\centering
\caption{Datasets}
\label{tab:datsets}
\fontsize{8}{10}\selectfont
\begin{tabular}{ l l l l l }
\toprule
Name & Embedding dimension & Corpus size & Query size & Source\\
\midrule
SIFT & 128 & 1000000 & 10000 & http://corpus-texmex.irisa.fr/\\
Glove-100 & 100 & 1183514 & 10000 & https://nlp.stanford.edu/projects/glove/\\
GIST & 960 & 1000000 & 1000 &  https://www.cs.cmu.edu/enron/\\
Crawl & 300 & 1989995& 10000 & https://commoncrawl.org/\\
Audio & 192 & 53387 & 200 & https://www.cs.princeton.edu/cass/demos.htm\\
Msong & 420 &  992272& 200 &  http://www.ifs.tuwien.ac.at/mir/msd/\\

\bottomrule
\end{tabular}
\end{table}

\begin{figure}[h]
\centering
  \includegraphics[scale=0.6]{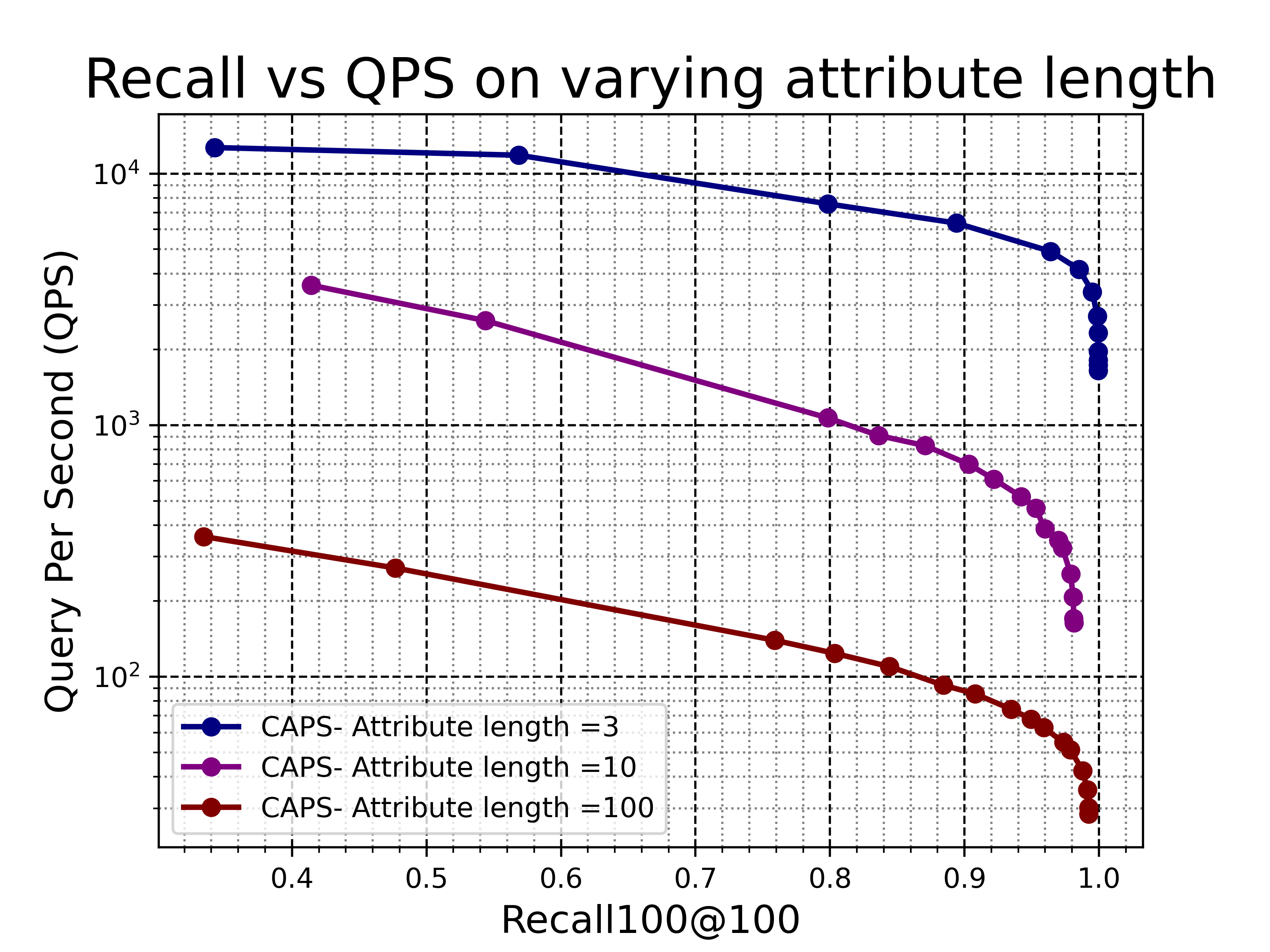}
  \caption{ Attribute length $L$ is 3, 10 and 100 for SIFT 1M dataset. The Attribute length is the number of attributes present for each item at indexing time. During the query, the user selects random attributes with probability 1, 0.3, and 0.03 for L =3,10,100, respectively. This represents the real search scenario where the user selects only a limited number of attribute constraints, but an item can have many attributes. }
  \label{fig:LargeAttr}
\end{figure}

\subsection{Experiment on Larger attribute lengths}
Following is the QPS vs Recall experiment on the Larger attribute length. We vary attribute length $L$ from 3 to 100 and notice the expected drop in query time with larger $L$.

\subsection{Query time complexity}
CAPS creates the index with a partition identification function $f(.)$, sub-partition lookup, and index files. The query time is the summation of top $m$ partition identification time + sub-partition identification time + brute force filter from the given sub-partition + brute force distance computation on the constrained satisfied items from the sub-partition. For the k-means this is 
$$ QT= Bd + B\log m + mO(1) + \sum_{i=1}^{m} \sum_{j=1}^{h+1}(\mathbb{I}(A(p)=b)|p_{i,j}|) (L + d \mathbb{I}(Attribute match) )$$
The Attribute match is the constrained match on query attribute $b$ and the item attributes of $p_{i,j}$. If the attributes are correlated with the embedding feature, we will observe the $\mathbb{I}(Attribute match)$ be 1 for most of the items in the top partitions match. However, in the worst case scenario, we can get $\mathbb{I}(Attribute match)$ to be uniformly distributed over all the item embeddings. Assuming this uniform distribution probability a.k.a sparsity to be $\gamma$, we can write
$$ QT \leq Bd + B\log m + mO(1) + \sum_{i=1}^{m} \sum_{j=1}^{h+1}(\mathbb{I}(A(p)=b)|p_{i,j}|) (L + d \gamma)$$

If the query specifies all the attributes, only one sub-partition is selected from each partition. Hence assuming a balanced sub-partitioning, we have
$$ QT \leq Bd + B\log m + mO(1) + \sum_{i=1}^{m} \frac{|P_i|}{h+1} (L + d\gamma )$$

Under an assumption that the number of partitions is polynomial in $N$,  $|B|= N^t$, where $t<1$, the sublinear query time is achieved in 
$$ QT \leq  N^td +  N^t\log m + m + m \frac{ N^{1-t}}{h+1} (L + d\gamma ) = QT_{UB}$$

The minimum of the query time upper bound is achieved when $d(QT_{UB})/dt = 0$ 
$$t = \frac{\log N \theta}{2 \log N}, \ \ \ \ \text{Where} \  \theta = \frac{(L+d\gamma)m}{(h+1)(d+\log m)}$$

\subsection{Index size}
The number of bytes used in RAM from index and data is NN model size + CSR sub-partition lookup + sub-partition key + data (including attributes + vectors)
$$Size(Index+ data)= 4Bd + 4N + 4B(h+1) + 2B(h+1)r + 4Nd + NLr$$ 
where $r = \log L \mod 8$ is the attribute precision. For example, for a total of 180 attribute values, we only need $r=1$, i.e. 1 byte for representation
$$Size(Index+ data)= B(4d + 2(h+1)(2+r)) + N(4d +1+ rL)$$


\end{document}